\documentclass[twocolumn,journal]{IEEEtran}
\usepackage{cite}
\usepackage{graphicx}
\usepackage{amsmath}
\usepackage{gettitlestring}
\begin{document}
\title{Optimal Spectrum Access for Cognitive Radios}
\author{Ahmed El Shafie\\
 \begin{tabular}{c}
Wireless Intelligent Networks Center (WINC), Nile University, Giza, Egypt. \\
\end{tabular}
}
\date{}
\maketitle
\begin{abstract}
In this paper, we investigate a time-slotted cognitive setting with buffered primary and secondary users. In order to alleviate the negative effects
of misdetection and false alarm probabilities, a novel design of spectrum access mechanism
is proposed. We propose two schemes. First, the SU senses primary channel to exploit the periods of silence, if the PU is declared to be idle, the SU randomly accesses the channel with some access probability $a_s$. Second, in addition to accessing the channel if the PU is idle, the SU possibly accesses the channel if it is declared to be busy with some access probability $b_s$.
The access probabilities as function
of the misdetection, false alarm and average primary arrival rate are obtained via solving an optimization
problem designed to maximize the secondary service rate given
a constraint on primary queue stability. In addition, we propose a variable sensing duration schemes where the SU optimizes over the optimal sensing time to achieve the maximum stable throughput of the network. The results reveal the performance gains of the proposed schemes over the conventional sensing scheme.
We propose a method to estimate the mean arrival rate and the outage probability of the PU based on the primary feedback channel, i.e., acknowledgments (ACKs) and negative-acknowledgments (NACKs) messages.

\end{abstract}
\section{Introduction}

The electromagnetic radio spectrum is a precious resource, the use of which by transmitters and receivers is licensed by governments \cite{haykin2005cognitive}. Regulatory bodies have come to realize and announce that most of the time, large portions of certain licensed frequency bands remain
unused. The intuitive intention behind secondary spectrum licensing is to efficiently increase the spectral
usage of the network, while, depending on the type of licensing, not perturbing
the primary users (PUs) (higher priority users). Cognitive radio (CR) systems are seen as a candidate prime solution that can significantly mitigate the current low spectral efficiency in the electromagnetic spectrum. A CR is defined as an intelligent wireless communication system that is fully aware of its environment and
uses methodologies of learning and reasoning in order to
dynamically adapt its transmission parameters (e.g., operating spectrum, modulation, and transmission power) to access portions of spectrum by exploiting the existence of spectrum holes left unused by a primary system.

 The CRs (secondary users (SUs)) exploit periods of silence of PUs under certain quality of service (QoS) for the PUs. In a typical cognitive radio setting the cognitive transmitter senses primary activity and decides on accessing the channel on the basis of the sensing outcome, which we refer to as conventional sensing scheme, $\mathcal{S}_{c}$. This approach is problematic because sensing may affect primary QoS. Spectrum sensing to detect the presence of the
PUs is, therefore, a fundamental requirement in cognitive
radio networks.

In a fixed frame size, the longer sensing time
will shorten the allowable data transmission time of the SU while improving the sensing
performance \cite{peh2009optimization}. Hence, a sensing-throughput tradeoff problem was formulated
in \cite{liang2008sensing} to find the optimal sensing time that maximizes the secondary
users' throughput while providing adequate protection to the
PU. Both the sensing time and the
cooperative sensing scheme affect the spectrum sensing performance,
such as the probabilities of detection and false alarm. These probabilities
affect the throughput of the secondary users since they determine
the reusability of frequency bands \cite{peh2009optimization}. In \cite{pei2009sensing}, the authors investigated the sensing-throughput
tradeoff problem for a multiple-channel CR
network. In particular, using the sensing-throughput tradeoff
metric, they designed the optimal spectrum sensing
time and power allocation schemes so as to maximize the
aggregate ergodic throughput of the CR network to
guarantee the QoS of the PUs without exceeding the power limit of the secondary transmitter. Recently, the authors of \cite{KarimSultan} proposed a random access scheme where the SU randomly accesses the primary channel with some access probability without employing any sensing scheme. The SU exploits the feedback messages of the PU.

In the case where the communicating terminals have queues, queue interaction renders difficult the analysis of stability and other relevant system characteristics such as the queueing delay \cite{rao1988stability,stabN}. The study of interacting queues has received much attention because of its natural existence in applications as well as its theoretical interest. The authors in \cite{tsy} provided a rigorous treatment of the problem and implicitly used the concept of a dominant system, which was explicitly introduced in \cite{rao1988stability}.

Queue stability within the context of cognitive radios has been investigated in many papers such as \cite{stability_cog1,stability_cog2,erph}. The authors of \cite{stability_cog1} studied a cognitive network with one primary and one secondary users. The secondary terminal adjusts its power such that the secondary queue mean service rate is maximized while preserving the stability of both the primary and secondary queues. In \cite{stability_cog2}, a cognitive scenario is considered with one primary node and multiple secondary nodes transmitting over a collision channel to a common receiver. All the secondary users attempt to sense and access the channel at the same time. The QoS specification employed in obtaining the secondary access probabilities is the average delay of primary packets, together with the stability of all queues.

In this work, we try to address the impact of sensing the electromagnetic spectrum for $\tau$ seconds of the time slot proceeded by randomly accessing it with some access probability based on the sensing outcome. We investigate the maximum stable throughput of the network. We optimize over the sensing duration that the SU can use to maximize its maximum stable throughput. In addition, we try to address when the SU can switch between sensing the channel for $\tau$ seconds of the time slot and randomly accesses it without employing any sensing schemes. The optimal access probabilities are function of the many parameters, including the sensing durations and the mean arrival rate of the PU. In order to obtain the mean arrival rate of the PU, we propose a method to estimate the mean arrival rate of the PU by exploiting the PU feedback messages, i.e., ACKs and NACKs. The theoretical and numerical results show that pairs of misdetection and false alarm probabilities may exist such that sensing the primary channel for very small duration overcomes sensing it for large portion of the time slot. In addition, for certain average arrival rate to the primary queue pairs of misdetection and false alarm probabilities may exist such that the random access without sensing overcomes the random access with long sensing duration.

\begin{figure}
\center
  \includegraphics[width=1\columnwidth]{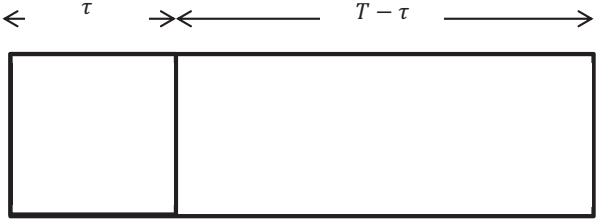}\\
  \caption{Time slot structure.}\label{fig20}
\end{figure}

The rest of the paper is organized as follows. Next we describe the system model adopted in this paper. The proposed scheme are discussed in Section \ref{sec3}. In Section \ref{sec5}, we provide some numerical results, and finally, we conclude the paper in Section \ref{sec6}.
\section{System Model}\label{sec2}
\subsection{MAC Layer}
\begin{figure}
  \includegraphics[width=1\columnwidth]{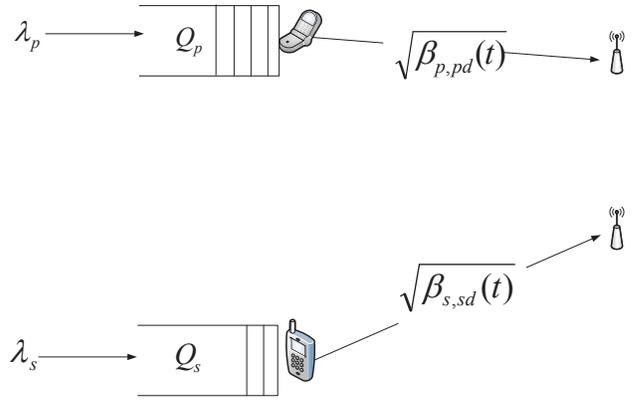}\\
  \caption{Primary and secondary links and queues.}\label{fig1}
\end{figure}

Our network consists of one PU and one SU as depicted in Fig. \ref{fig1}. The SU senses the primary channel to detect the possible activities of the PU. The main contribution in this paper is that the SU randomly accesses the channel preceded by spectrum sensing for $\tau$ seconds of the time slot instead of accessing with probability one. First, we consider the case where the SU randomly accesses the primary channel only if the PU is declared to be idle. Secondly, we investigate the case where the SU randomly accesses the channel if the PU is sensed to be idle with access probability $a_s$ and if it is declared to be busy with probability $b_s$. This is because the SU tries to mitigate the impact of misdetection and false alarm probabilities in order to maximize its throughput. Fig\ \ref{fig20} shows the frame structure designed for a cognitive
radio network with periodic spectrum sensing where each
frame consists of one sensing slot and one data transmission
slot.

The probability that the SU misdetects the primary activity is $P_{\rm MD}$ and the probability that the SU's sensor generates false alarm is denoted as $P_{\rm FA}$. We assume that the primary transmitter has a buffer $Q_p$
to store the incoming traffic packets, while the secondary
transmitter has a buffer $Q_s$ to store its
own arrived traffic packets. We assume all buffers are of
infinite length. We consider time-slotted transmission where all packets have the same size and one time slot is sufficient for the transmission
of a single packet. The arrival processes of the primary and the secondary transmitters are assumed to be independent Bernoulli processes with mean arrival rates $\lambda_p$ and $\lambda_s$ packet per time slot, respectively.

\subsection{Physical Layer}
%, where $\overline{P}_{j,d}$ denotes the probability of correct reception of the packet transmitted by user $j$ to its respective receiver $d$.
In this work, we characterize the success and failure of packet
reception by outage events and outage probability. The probability of outage event of the link between SU and secondary destination (SD), $P_{s,sd}$, can be calculated as in \cite{ShafieSultan}. The transmitter adjusts its transmission rate depending on when it starts transmission during the time slot.
Assuming that the number of bits in a packet is $b$ and the time slot duration is $T$, the transmission rate is
\begin{equation}
r=\frac{b}{T-\tau}.
\label{r_i}
\end{equation}
\noindent If transmission is preceded by a spectrum sensing period of $\tau$ units of time the time remaining for SU transmission is $T-\tau$. Consider the SU and its destination PD, an outage for the transmission occurs when the transmission rate exceeds channel capacity
\begin{equation}
P^{\left(\tau\right)}_{s,sd}={\rm Pr}\biggr \{r > W \log_{2}\left(1+\gamma_{s,sd} \beta_{s,sd}\right)\biggr\}
\end{equation}
\noindent where $\tau\in[0,T]$, $W$ is the bandwidth of the channel, $\gamma_{s,sd}$ is the received SNR when the channel gain is equal to unity, and $\beta_{s,sd}$ is the channel gain, which is exponentially distributed in the case of Rayleigh fading. The outage probability can be written as
\begin{equation}
P^{\left(\tau\right)}_{s,sd}={\rm Pr}\Big\{\beta_{s,sd}<\frac{2^{\frac{r}{W}}-1}{\gamma_{s,sd}}\Big\}
\end{equation}
\noindent Assuming that the mean value of $\beta_{s,sd}$ is $\sigma^2_{s,sd}$,
\begin{equation}
P^{\left(\tau\right)}_{s,sd}=1-\exp\bigg(-\frac{2^{\frac{r}{W}}-1}{\gamma_{s,sd}\sigma^2_{s,sd}}\bigg)
\end{equation}
\noindent Let $\overline{P}_{s,sd}=1-P_{s,sd}$\footnote{Throughout the paper $\overline{x}=1-x$.} be the probability of correct reception. It is therefore given by
\begin{equation}\label{correctreception}
\overline{P}^{\left(\tau\right)}_{s,sd}=\exp\bigg(-\frac{2^{\frac{b}{TW\left(1-\frac{\tau}{T}\right)}}-1}{\gamma_{s,sd}\sigma^2_{s,sd}}\bigg)
\end{equation}
Note that the outage probability increases as $\tau$ increases. The probability of channel outage of the link between the PU and the primary destination (PD), $P_{p,pd}$, has a similar formula with $\tau=0$ and the associated parameters.
\begin{equation}\label{correctreception}
\overline{P}_{p,pd}=\exp\bigg(-\frac{2^{\frac{b}{TW}}-1}{\gamma_{p,pd}\sigma^2_{p,pd}}\bigg)
\end{equation}
\subsection{Misdetection and False Alarm Probabilities}
In this paper, we adopt the formulas of misdetection and false alarm depicted in \cite{liang2008sensing}. If we choose the detection threshold as $\epsilon$, the probability
of false alarm is then given by

\begin{equation}
\begin{split}
P_{\rm FA}(\epsilon,\tau)=Q\biggr(\bigg(\frac{\epsilon}{\sigma^2_u}-1\bigg)\sqrt{\tau f_s}\biggr)
 \end{split}
\end{equation}
The probability of misdetection is given by
\begin{equation}
\begin{split}
P_{\rm MD}(\epsilon,\tau)=1-Q\biggr(\bigg(\frac{\epsilon}{\sigma^2_u}-\gamma-1\bigg)\sqrt{\frac{\tau f_s}{2\gamma+1}}\biggr)
 \end{split}
\end{equation}
where $\sigma_u^2$ is the additive Gaussian noise variance, $\gamma$ is the received SNR, $f_s$ the sampling frequency, and $Q(.)$ is the complementary distribution function of the standard Gaussian.

For a target false alarm, $P_{\rm FA}$, the probability of misdetection is given by:
\begin{equation}
\begin{split}
P^{\left(\tau\right)}_{\rm MD}=1-P^{\left(\tau\right)}_{\rm D}=1-Q\bigg(\frac{1}{\sqrt{2 \gamma+1}}\big(Q^{-1}(P_{\rm FA})-\sqrt{\tau f_s}\gamma\big)\bigg)
 \end{split}
\end{equation}
Therefore, for a target probability of detection $P_{\rm D}$, the probability
of false alarm is related to the target detection probability
as follows:
\begin{equation}
\label{pmdpfa}
\begin{split}
P^{\left(\tau\right)}_{\rm FA}=Q\bigg(\sqrt{2 \gamma+1}Q^{-1}(1-P_{\rm MD})+\sqrt{\tau f_s}\gamma\bigg)
 \end{split}
\end{equation}
where $P_{\rm MD}$ is the target probability of misdetection with which the
PU is defined as being sufficiently protected. In
practice, the target probability of misdetection $P_{\rm MD}$ is chosen to
be close to but greater than zero, especially for low SNR regime. If the misdetection probability equals to zero, then the SU cannot use the primary channel.

Obviously, for a given time slot duration $T$, the longer the
sensing time $\tau$, the shorter the available data transmission
time $(T - \tau )$. This reduction in the data time is interpreted as increasing in the probability of channel outage. On the other hand, from (\ref{pmdpfa}), since Q(z) is a monotonically decreasing function of $z$, for a given target
probability of detection, $P_{\rm D}$, the longer the sensing time, the
lower the probability of false alarm, which corresponds to the
case that the secondary network can use the channel with a
higher chance. The objective of sensing-throughput tradeoff is
to identify the optimal sensing duration $\tau$ for the SU
that the achievable throughput of the secondary network is
maximized while the PU is sufficiently protected. This protection is interpreted as stability of the primary queue.
Mathematically, the optimization problem can be stated as
\begin{equation}
\begin{split}
\max_{\Gamma} \,\,\ \lambda_s&=\mu_s\\ {\rm s.t.} \,\,\  \mu_p&\ge\lambda_p
\end{split}
\end{equation}
where $\Gamma$ is the set of optimization variables of the system.

For simplicity of presentation, we omit the superscripts of the notations, i.e., $P^{\left(\tau\right)}_{\rm MD}$, $P^{\left(\tau\right)}_{\rm FA}$ and $\overline{P}^{\left(\tau\right)}_{s,sd}=\overline{P}_{s,sd}$.

Note that the calculation of the access probability requires a previous knowledge of the mean arrival rate of the PU, i.e., $\lambda_p$, and the outage probability of the primary link, i.e., $P_{p,pd}$, which can be \textbf{estimated} easily, i.e., by overhearing the feedback channel and counting the total number of acknowledgments (ACKs) and negative-acknowledgments (NACKs). Next we discuss the estimation process and errors in estimation and errors in feedback message detection.
\subsection{Learning Phase (LP) and Regular Phase (RP)}
\begin{figure}
  \includegraphics[width=1\columnwidth]{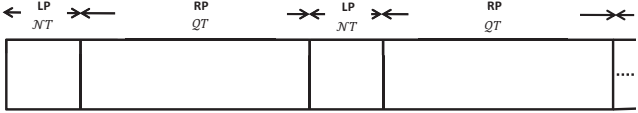}\\
  \caption{System operation.}\label{fig1000}
\end{figure}

: 
Assume that the operation of the SU consists of two phases; learning phase (LP) for $\mathcal{N}T$ seconds to estimate the required parameters from the environment and regular phase (RP) for $\mathcal{Q}T$ seconds to access the channel according to one of the proposed schemes, where $\mathcal{Q}\gg \mathcal{N}$ (see Fig. \ \ref{fig1000}) and $\mathcal{N}$ should be long enough for accurate estimations. The estimation of the arrival rate can be done by counting the total number of ACKs and NACKs during some predefined period of time, i.e., during $\mathcal{N}T$ seconds where $\mathcal{N}$ is the total number of time slots used for learning. The mean service rate can be estimated by observing the ratio of ACKs to total number of transmissions during LP. The ratio of total transmissions to the total number of slots gives the probability of the queue being nonempty, $\frac{\lambda^{\left(\rm est\right)}_p}{\mu^{\left(\rm est\right)}_p}$. Whereas, the total number of ACKs over the total number of ACKs and NACKs provides the probability of correct reception of a primary packet, i.e., $\mu^{\left(\rm est\right)}_p$.\footnote{ It is an estimation of the mean service rate of the PU in case of silent SU.} Mathematically, the estimated mean arrival rate, $\lambda^{\left(\rm est\right)}_p$, is given by:

\begin{equation}
\begin{split}
\lambda^{\left(\rm est\right)}_p&=\frac{\mathcal{A}}{\mathcal{N}}\\&=\frac{\hbox{Total number of ACKs during LP}}{\hbox{Total number of time slots during LP}}  \\& \,\,\,\,\,\,\,\,\,\,\,\,\,\,\,\,\,\,\,\,\,\,\,\,\,\,\,\,\,\,\,\,\,\,\,\,\,\,\,\,\,\,\,\,\,\,\,\,\,\,\,\,\,\ \hbox{\ \ packets per time slot}
\end{split}
\end{equation}
 The ACK/NACK messages are very short compared to the duration of the time slot and are always received correctly due to the use of strong channel codes. A correctly received packet is removed from the respective transmitter's queue. Note that, even if the feedback messages are received erroneously with some probability of error $P_e$\footnote{Note that we assume that if the feedback packet is received properly, i.e., without errors, then it is counted, otherwise the SU considers the PU is idle in the previous time slot and the counter adds zero.}, we still can estimate the mean arrival rate of the PU. In this case, the estimated mean arrival rate of the PU is given by
 \begin{equation}
\begin{split}
&\lambda^{\left(\rm est\right)}_p=\frac{\mathcal{A}(1-P_e)}{\mathcal{N}}
 \end{split}
\end{equation}
 where $\mathcal{A}$ is the total number of decoded ACKs, i.e., when $P_e=0$ and $\mathcal{M}(1-P_e)$ is the total number of correctly overheard ACKs during LP

In order to protect the PU from errors in the mean arrival rate estimation, we can add some protection on the stability constraint of the PU, i.e., adding more restriction on Loynes's law, the extra term equals to the negative of the maximum average estimation error of the mean of the estimator. In this case the optimization problem is given by
\begin{equation}
\begin{split}
\max_{\Gamma} \,\,\ \lambda_s&=\mu_s\\ {\rm s.t.} \,\,\  \mu_p&\ge\lambda_p+\mu_{\rm pe}
\end{split}
\end{equation}
where $\mu_{\rm pe}$ is a nonnegative constant represents more protection for the PU. This protection is more than the required for stability. Note that this increasing in the mean service rate of the PU is interpreted as decreasing in the tolerable primary delay, i.e., if the estimation process is error-free. Assume that the expected value of the mean arrival rate estimation error $\mathcal{E}\bigg\{\lambda_p-\lambda^{\left(\rm est\right)}_p\bigg\}$ is a \textbf{random} variable varies over the set $[-e_{\lambda_p},e_{\lambda_p}]$ where $e_{\lambda_p}\ge 0$. The primary delay is given by $D_p=\frac{1-\lambda_p}{\mu_p-\lambda_p}\le \infty$ \cite{gambini2007stability,sadek}. In case we designed the allowable mean service rate of the primary queue to be greater than $\lambda_p+\mu_{\rm pe}$, i.e., $\mu_p\ge\lambda_p+\mu_{\rm pe}$ then the designed delay is given by $D_p\le \frac{1-\lambda_p}{\mu_{\rm pe}}$. If the primary mean arrival rate is estimated perfectly, error-free, then the primary delay will be lower than $\frac{1-\lambda_p}{\mu_{\rm pe}}$. On the other hand, if there is some errors, say $\hat{\lambda}_p$, then the delay is given by $\frac{1-\lambda_p}{\mu_{\rm pe}-\hat{\lambda}_p}$, if $\hat{\lambda}_p=e_{\lambda_p}\ge0$, then the primary delay, under the usage of $\mu_{\rm pe}$, is greater the perfect estimated case, but still less than $D_p=\infty$ at stability boundaries. On the other hand, if $\hat{\lambda}_p=-e_{\lambda_p}<0$ the primary node is totally protected (more than expected or designed) and the delay is less than the perfect estimation case. Thus, it is better to design the system based on the \textbf{maximum positive} error of the mean arrival rate estimator to guarantee full protection for the PU, i.e., $\mu_{\rm pe}\ge e_{\lambda_p}$. However, the estimation of the primary mean arrival rate is out of the scope of this paper.

The estimation of the outage probability of the primary link, i.e., the link between the primary transmitter and receiver, can be done as follows. The probability of \textbf{complement outage} is obtained by dividing the total number of ACKs during the learning phase over the total number of the overheard ACKs and NACKs during the learning phase. Mathematically, the probability of \textbf{correct} reception of a primary packet is given by
 \begin{equation}
 \label{1111}
\begin{split}
\overline{P}_{p,pd}^{\left(\rm est\right)}=\frac{\mathcal{A}}{\mathcal{M}}
 \end{split}
\end{equation}
 In case of errors in the primary feedback message detection, the denominator and numerator of (\ref{1111}) are multiplied by $(1-P_e)$, thus $\overline{P}_{p,pd}^{\left(\rm est\right)}$ is independent of $P_e$.
\subsection{Stability Analysis}
 Let us denote the queue sizes of the transmitting terminals at any time instant $t$ by $Q_i^t$. Then, $Q_i^t$ evolves according to
\begin{equation}\label{queue}
    Q_i^{t+1}=\bigr(Q_i^t-\mathcal{U}_i^t\bigr)^{+}+\mathcal{A}^t_i
\end{equation}
where $\mathcal{U}_i^t$ is the number of departures in time slot $t$. $\mathcal{A}^t_i$ denotes the number of arrivals in time slot $t$ and is a stationary process by assumption with finite mean $\mathcal{E}\{\mathcal{A}^t_i\}=\lambda_i$. The function $(.)^{+}$ is defined as $(x)^{+}=\max(x,0)$ and $\mathcal{E}\{.\}$ denotes the expected value. We assume that
departures occur before arrivals, and the queue size is measured at the beginning of the time slot \cite{sadek}.

A fundamental performance measure of a communication network is the stability of its queues. We are interested in the queues size. More rigourously, stability can be defined as follows \cite{szpankowski1994stability,sadek}.

\emph{Definition:} Queue $i\in\{p,s\}$ is stable, if
\begin{equation}\label{stabilityeqn}
   \lim_{t \rightarrow \infty  }{\rm Pr}\{Q_i^t<y\}=F(y) \hbox{ and}  \lim_{y \rightarrow \infty} F(y)=1
\end{equation}
If the arrival and service processes are strictly stationary, then we can apply Loynes's theorem to check for stability conditions \cite{loynes1962stability,sadek}. This theorem states that if the arrival process and the service process of a queue are strictly stationary processes, and the average service rate is greater than the average arrival rate of the queue, then the queue is stable, otherwise the queue is unstable.

To study the stability region of the network, we
note first that, since the secondary and primary queues are
interacting. In other words, the service rate of a given queue
is dependent on the state of other queue, i.e., whether
it is empty or not. Studying the stability conditions
for interacting queues is a difficult problem that has been
addressed for ALOHA systems \cite{rao1988stability,stabN}. The concept of
dominant systems was introduced and employed in \cite{rao1988stability} to
help find bounds on the stability region of ALOHA with
collision channel. The dominant system in \cite{rao1988stability} was defined
by allowing a set of terminals with no packets to transmit
to continue transmitting dummy packets. In this manner, the
queues in the dominant system stochastically dominate the
queues in the original system. Or in other words, with the
same initial conditions for queue sizes in both the original and
dominant systems, the queue sizes in the dominant system are
not smaller than those in the original system.

To study the stability of the interacting system of queues
consisting of secondary and primary queues, we make use of the
dominant system approach to decouple the interaction between
queues. We define the dominant system as follows
\begin{itemize}
\item Arrivals at each queue in the dominant system are the
same as in the original system.
\item Time slots assigned to primary node are identical
in both systems.
\item The outcomes of the ``coin tossing" (that determines
transmission attempts of secondary node) in
every slot are the same.
\item Channel realizations for both systems are identical.
\item The noise generated at the receiving ends of both systems
is identical.
\item In the dominant system, secondary node attempts to
transmit dummy packets when its queue is empty.

For a proof that the dominant system's stability conditions are
necessary and sufficient for the stability of the original system
see Appendix B.

\end{itemize}

%Due to queues interaction we consider the case of backlogged SU where the SU sends dummy packets when its queue empties(the SU always has packets to send). Note that this system is a lower bound on the original system because we consider that the SU interferes with the PU each time slot.
\section{Proposed Schemes}\label{sec3}
\subsection{Conventional Spectrum Sensing $\mathcal{S}_c$}
In a conventional spectrum sensing scheme, the SU senses the channel for $\tau$ seconds from the beginning of the time slot to detect the possible activities of the PU. If the PU is sensed to be idle the SU transmit the packet at the head of its queue with probability one. If the channel is declared to be idle the SU does not transmit. The primary queue is served if the SU correctly detects its activity and the link between PU and PD is not in outage. While the secondary queue is served if the primary queue is empty, the channel between the SU and its respective receiver is not in outage. Note that the probability that the PU being empty is given by
\begin{equation}
\begin{split}
{\rm Pr}\bigg\{Q_p=0\bigg\}=1-\frac{\lambda_p}{\mu_p}
\end{split}
\end{equation}
Thus, the average service rates of the nodes in this system are given by:
\begin{equation}
\begin{split}
\mu_p&=\overline{P}_{p,pd}\overline{P}_{MD}
 \end{split}
\end{equation}
\begin{equation}
\begin{split}
\mu_s&=\overline{P}_{s,sd}\overline{P}_{FA} \bigg(1-\frac{\lambda_p}{\overline{P}_{p,pd}\overline{P}_{MD}}\bigg)
\end{split}
\end{equation}
The maximum stable throughput, for specific $\tau$, is given by:
\begin{equation}
\begin{split}
\mathcal{R}(\mathcal{S}_c|\tau)&\!=\!\biggr\{\!(\lambda_p,\lambda_s)\!:\!\lambda_s \!<\!\overline{P}_{s,sd}\overline{P}_{FA} \bigg(\!1\!-\!\frac{\lambda_p}{\overline{P}_{p,pd}\overline{P}_{MD}}\!\bigg)\!\biggr\}
\end{split}
\end{equation}

The stability region of a backlogged SU is given by:
\begin{equation}
\begin{split}
\mathcal{R}(\mathcal{S}_c)&=\bigcup_\tau \mathcal{R}(\mathcal{S}_c|\tau)
\end{split}
\end{equation}
According to  Loynes's theorem the condition on stability of the PU and the SU are given by:
\begin{equation}
\begin{split}
\lambda_i <\mu_i, \ \ \hbox{and $i\in \{p,s\}$}
\end{split}
\end{equation}
The union over all possible values of $\tau$, $\bigcup_\tau \mathcal{R}(\mathcal{S}_c|\tau)$, can be obtained by solving the following optimization problem:
\begin{equation}
\begin{split}
\max_{\tau} \,\,\ \lambda_s&=\mu_s\\ {\rm s.t.} \,\,\ \lambda_p &\le \mu_p
\end{split}
\end{equation}
Thus,
\begin{equation}\label{100}
\begin{split}
& \max_{\tau} \,\,
\lambda_s=\overline{P}_{s,sd}\overline{P}_{FA} \bigg(1-\frac{\lambda_p}{\overline{P}_{p,pd}\overline{P}_{MD}}\bigg)\\
&\,\,{\rm s.t.} \,\,\,\,\  0 \le \frac{\tau}{T} \le 1,\ \\& \,\,\,\,\,\,\,\,\,\,\,\,\ \lambda_p\le\overline{P}_{p,pd}\overline{P}_{MD}
\end{split}
\end{equation}
Note that for a target $P_{\rm FA}$, both $\overline{P}_{s,sd}$ and probability of misdetection $P_{\rm MD}$ are functions of $\tau$.
\subsection{First Proposed Random Access Scheme $\mathcal{S}_1$}
In this subsection, we assume that the SU randomly accesses the channel if and only if the PU is declared to be idle. For the system with a backlogged SU, denoted as $\mathcal{S}_1$, a packet from the PU is served if the complement of the event that the SU detects primary transmission correctly and accesses the channel is true and the channel between PU and PD is not in outage. The average service rate of the PU is given by:
\begin{equation}
\begin{split}
\mu_p&=\overline{P}_{p,pd}\bigg(1-a_{s}P_{\rm MD}\bigg)
 \end{split}
\end{equation}
 Now consider the secondary queue. Given that the primary queue is empty, a packet from $Q_s$ is served if the SU detects the primary activity correctly, it decides to access the channel, and the channel between SU and SD is not in outage. Thus, the SU average service rate is given by:
\begin{equation}
\begin{split}
\mu_s&=a_s \overline{P}_{s,sd}\overline{P}_{FA} \bigg(1-\frac{\lambda_p}{\overline{P}_{p,pd}\bigg(1-a_{s}P_{\rm MD}\bigg)}\bigg)
\end{split}
\end{equation}
One method to characterize the closure of the rates pair $(\lambda_p,\lambda_s)$, to obtain the stability region, is to solve a constrained
optimization problem to find the maximum feasible $\lambda_s$ corresponding
to each feasible $\lambda_p$ as $a_s$ varies over $[0,1]$ and $\tau$ over $[0,1]$. For a fixed $\lambda_p$, the maximum
stable arrival rate for the secondary queue is given by solving the following optimization problem \cite{sadek}:
\begin{equation}
\begin{split}
& \max_{a_s,\tau} \,\,
\lambda_s=a_s \overline{P}_{s,sd}\overline{P}_{FA} \bigg(1-\frac{\lambda_p}{\overline{P}_{p,pd}\bigg(1-a_{s}P_{\rm MD}\bigg)}\bigg)\\
&\,\,{\rm s.t.} \,\,\,\,\  0 \le a_s,\frac{\tau}{T} \le 1,\ \ \lambda_p\le\overline{P}_{p,pd}\bigg(1-a_{s}P_{\rm MD}\bigg)
\end{split}
\end{equation}
For a fixed $\tau$, the optimization problem is \textbf{concave} and it can be readily solved using Lagrangian multipliers. For the problem to be feasible the average primary arrival rate $\lambda_p$ must be less than or equal to $\overline{P}_{p,pd}$. The optimal access probability is given by:
\begin{equation}
\begin{split}
a^*_s=\max\biggr(\min\biggr(\frac{1-\sqrt{\frac{\lambda_p}{\overline{P}_{p,pd}}}}{P_{\rm MD}},1\biggr),0\biggr)
\end{split}
\end{equation}
The stability region of a backlogged SU for a fixed $\tau$ is given by:
\begin{equation}
\begin{split}
\mathcal{R}(\mathcal{S}_1|\tau)&\!=\!\biggr\{\!(\!\lambda_p,\lambda_s\!)\!:\!\lambda_s \!<\!a^*_s \overline{P}_{s,sd}\overline{P}_{\rm FA} \!\bigg(\!1\!-\!\frac{\lambda_p}{\overline{P}_{p,pd}\bigg(1\!-\!a^*_{s}P_{\rm MD}\!\bigg)\!}\!\bigg)\!\biggr\}\!
\end{split}\label{e1}
\end{equation}
with $\lambda_p\le \overline{P}_{p,pd}$.

The stability region of a backlogged SU is given by:
\begin{equation}
\begin{split}
\mathcal{R}(\mathcal{S}_1)&=\bigcup_\tau \mathcal{R}(\mathcal{S}_1|\tau)
\end{split}
\end{equation}
 Note that the optimal access probability $a^*_s$
guarantees the stability of the PU queue which requires the access probability to satisfy the stability condition of the primary queue, i.e., $\lambda_p\le\mu_p=\overline{P}_{p,pd}\bigg(1-a^*_{s}P_{\rm MD}\bigg)$.

\subsection{Second Proposed Random Access Scheme $\mathcal{S}_2$}
 In addition to the operation of the SU in the first proposed scheme, the SU randomly accesses the channel even if the PU is declared to be busy with some access probability $b_s$, this scheme is denoted as $\mathcal{S}_2$. This is useful to mitigate the impact of false alarm probability.  Given that the channel between PU and PD is not in outage, a packet from the primary queue $Q_p$ can be served in either one of the following events: 1) if the SU detects the primary activity correctly and decides not to access the channel (which happens with probability $\overline{a}_s$); or 2) if the SU misdetects the primary activity and decides not to access the channel (which happens with probability $\overline{b}_s$). The average service rate of the primary queue can be given by:
\begin{equation}
\begin{split}
\mu_p&\!=\! P_{\rm MD}\overline{a}_{s}\overline{P}_{p,pd}\!+\! \overline{P}_{\rm MD}\overline{b}_{s}\overline{P}_{p,pd}
\end{split}
\end{equation}
Given that the primary queue is empty, a packet from $Q_s$ is served in either one of the following events: 1) if the SU detects the primary activity correctly, it decides to access the channel, and the channel between SU and SD is not in outage; or 2) if SU's sensor generates false alarm, the SU decides to access the channel with probability $b_s$ and the channel between SU and SD is not in outage. The average service rate of the secondary queue can be given by:
\begin{equation}
\begin{split}
\mu_s&=\biggr[a_s \overline{P}_{s,sd}\overline{P}_{\rm FA}+ b_s \overline{P}_{s,sd}P_{\rm FA} \biggr] \bigg(1-\frac{\lambda_p}{\mu_p}\bigg)
\end{split}
\end{equation}
The maximum stable throughput is given by solving the following optimization problem:
\begin{equation}\label{200}
\begin{split}
& \max_{a_s,b_s,\tau} \,\,
\lambda_s=\biggr[a_s \overline{P}_{s,sd}\overline{P}_{\rm FA}+ b_s \overline{P}_{s,sd}P_{\rm FA} \biggr] \bigg(1-\frac{\lambda_p}{\mu_p}\bigg)\\
& \ \ \ {\rm s.t.} \  0 \! \le\! a_s,b_s,\frac{\tau}{T}\! \le \!1 \\& \,\,\,\,\,\,\,\,\,\,\,\,\,\,\,\ \lambda_p\! \le \! P_{\rm MD}\overline{a}_{s}\overline{P}_{p,pd}\!+\! \overline{P}_{\rm MD}\overline{b}_{s}\overline{P}_{p,pd}
\end{split}
\end{equation}
The problem can be reduced to
\begin{equation}
\begin{split}
&\max_{\mathcal{T},\tau} \ \ \mathcal{C}^\dagger \mathcal{T}+\lambda_p \frac{\mathcal{C}^\dagger \mathcal{T}}{\mathcal{D}^\dagger \mathcal{T}+\mathcal{F}}\\
&\,\,{\rm s.t.} \,\,\,\,\  0 \le a_s,b_s,\frac{\tau}{T} \le 1\ \\& \,\,\,\,\,\,\,\,\ \mathcal{D}^\dagger \mathcal{T}+\mathcal{F}\le 0
\label{form}
\end{split}
\end{equation}
 Where $\dagger$ denotes vector transposition, $\mathcal{F}\!=\! \lambda_p\!-\!\overline{P}_{p,pd}$, and
\begin{equation}
\begin{split}
  \mathcal{D}\! & =\!\biggr[\!\begin{array}{c}P_{\rm MD}  \overline{P}_{p,pd}\\ \overline{P}_{\rm MD} \overline{P}_{p,pd} \\ \end{array}\!\biggr] \ \\ \mathcal{T}\!&=\! \biggr[\!\begin{array}{c}
a_s\\
b_s\end{array}\!\biggr] \\ \ \mathcal{C}\!&=\!\biggr[\!\begin{array}{c}
\overline{P}_{s,sd}\overline{P}_{FA}\\
\overline{P}_{s,sd}P_{\rm FA}\end{array}\!\biggr]
 \end{split}
\end{equation}
Fixing $b_s$ and $\tau$, the optimization problem can be stated as
\begin{equation}
\begin{split}
& \max_{a_s} \,\,
a_s \overline{P}_{\rm FA}\!+\!\frac{\lambda_p}{\overline{P}_{p,pd}} \frac{a_s \overline{P}_{\rm FA}\!+\!b_s P_{\rm FA} }{a_sP_{\rm MD}\!-\! (\overline{P}_{\rm MD}\overline{b}_{s}\!+\!P_{\rm MD})}\\
&\,\,{\rm s.t.} \,\,\,\,\  0 \le a_s\le 1 \\ & \,\,\,\,\,\,\,\,\,\,\,\,\,\,\,\,\  a_s\le \frac{P_{\rm MD}+ \overline{P}_{\rm MD}\overline{b}_{s}-\frac{\lambda_p}{\overline{P}_{p,pd}}}{P_{\rm MD}}
\end{split}
\end{equation}
The optimization problem, given $b_s$ and $\tau$, is \textbf{concave} and can be readily solved. The solution is provided in Appendix A. From Appendix A, for fixed $b_s$ and $\tau$ the problem is feasible if $P_{\rm MD}\!+\! \overline{P}_{\rm MD}\overline{b}_{s}\!\ge\! \frac{\lambda_p}{\overline{P}_{p,pd}}$. The optimal value of $a_s$ is depicted in (\ref{equation_long}) at the top
of the following page.

%
%\begin{equation}\label{equation_long}
%\begin{split}
%a_s^*&\!=\! \max\biggr\{\!\min\biggr\{\! \frac{\bigg(P_{\rm MD}\!+\! \overline{P}_{\rm MD}\overline{b}_{s}\bigg)\!-\! \sqrt{\!\frac{\frac{\overline{P}_{\rm FA}\lambda_p}{\overline{P}_{p,pd}}\bigg(P_{\rm MD}\!+\! \overline{P}_{\rm MD}\overline{b}_{s}\bigg)\!+\!P_{\rm MD}\frac{P_{\rm FA}\lambda_p}{\overline{P}_{p,pd}}}{\overline{P}_{\rm FA}}}}{P_{\rm MD}}\!,\!\frac{P_{\rm MD}\!+\! \overline{P}_{\rm MD}\overline{b}_{s}\!-\!\frac{\lambda_p}{\overline{P}_{p,pd}}}{P_{\rm MD}}\!,\!1\biggr\},0\!\biggr\}
%%,\  \hbox{ with $ad \ge cf$}
%\end{split}
%\end{equation}

\begin{figure*}[!t]
\normalsize
\setcounter{equation}{37}
\begin{equation}\label{equation_long}
\begin{split}
a_s^*&\!=\! \max\biggr\{\min\biggr\{\! \frac{\bigg(P_{\rm MD}+ \overline{P}_{\rm MD}\overline{b}_{s}\bigg)\!-\! \sqrt{\!\frac{\frac{\overline{P}_{\rm FA}\lambda_p}{\overline{P}_{p,pd}}\bigg(P_{\rm MD}+ \overline{P}_{\rm MD}\overline{b}_{s}\bigg)\!+\!P_{\rm MD}\frac{P_{\rm FA}\lambda_p}{\overline{P}_{p,pd}}}{\overline{P}_{\rm FA}}}}{P_{\rm MD}},\frac{P_{\rm MD}+ \overline{P}_{\rm MD}\overline{b}_{s}-\frac{\lambda_p}{\overline{P}_{p,pd}}}{P_{\rm MD}},1\biggr\},0\biggr\}
%,\  \hbox{ with $ad \ge cf$}
\end{split}
\end{equation}
\hrulefill
\vspace*{1pt}
\end{figure*}

The maximum stable throughput for a fixed $\tau$ is given by:
\begin{equation}
\begin{split}
\mathcal{R}(\mathcal{S}_2|\tau)&=\biggr\{(\lambda_p,\lambda_s):\lambda_s <\mathcal{C}^\dagger \mathcal{T}^*+\lambda_p \frac{\mathcal{C}^\dagger \mathcal{T}^*}{\mathcal{D} \mathcal{T}^*+\mathcal{F}}\biggr\}
\end{split}\label{eqn5}
\end{equation}
The maximum stable throughput of $\mathcal{S}_2$ is given by the union over all possible values of $\tau$
\begin{equation}
\begin{split}
\mathcal{R}(\mathcal{S}_2)&=\bigcup_\tau \mathcal{R}(\mathcal{S}_2|\tau)
\end{split}
\end{equation}
Note that $\mathcal{S}_1\bigcup \mathcal{S}_2=\mathcal{S}_2$ because $\mathcal{S}_1$ is achieved from $\mathcal{S}_2$ by setting $b_s=0$.
\subsection{Random Access without Sensing Scheme $\mathcal{S}_\circ$}
In this system, denoted as $\mathcal{S}_\circ$, the SU accesses primary channel without employing any sensing schemes. The average primary and secondary service rates are given by:
\begin{equation}
\begin{split}
\mu_p&= \overline{a}_{s}\overline{P}_{p,pd} \\ \mu_s&=a_s \overline{P}_{s,sd} \bigg(1-\frac{\lambda_p}{\mu_p}\bigg)
\end{split}
\end{equation}
As in \cite{KarimSultan} the maximum stable throughput of a backlogged SU, after including the channels outage probability, is given by:
\begin{equation}
\begin{split}
\mathcal{R}(\mathcal{S}_\circ)&=\biggr\{(\lambda_p,\lambda_s):\lambda_s < \overline{P}_{s,sd}\bigg(1-\sqrt{\frac{\lambda_p}{\overline{P}_{p,pd}}}\bigg)^2\biggr\}
\end{split}\label{e3}
\end{equation}
Note that in this scheme the probability of the secondary channel outage is less than the outage probabilities in case of $\mathcal{S}_1$ and $\mathcal{S}_2$ because the SU does not waste $\tau$ seconds in sensing, i.e., $P_{s,sd}|_{\mathcal{S}_\circ}<P_{s,sd}|_{\mathcal{S}_j}$ and $j \in\{1,2\}$.

It should be noticed that given specific $\lambda_p$ and $\tau$, we possibly can find certain values of false alarm and misdetection probabilities such that when the SU randomly accesses the channel without sensing is better than randomly accessing preceded by sensing it for $\tau$ seconds and vice versa. The pair $(P_{\rm FA},P_{\rm MD})$ can be fully specified by the boundary points of the stability region of the schemes, i.e., using Eqns. (\ref{e1}), (\ref{eqn5}), and (\ref{e3}). In other words, the pair exists if and only if it satisfies the condition
 \begin{equation}
\begin{split}
\lambda_s|_{\tau,\lambda_p}\big(\mathcal{S}_2)<\lambda_s|_{\lambda_p}\big(\mathcal{S}_\circ\big)
\end{split}
\end{equation}
In addition, for a fixed $\tau_1$, one possibly find $\tau_2>\tau_1$ for each $\lambda_p$, such that
 \begin{equation}
\begin{split}
\lambda_s|_{\tau_1,\lambda_p}\big(\mathcal{S}_2)>\lambda_s|_{\tau_2,\lambda_p}\big(\mathcal{S}_2)
\end{split}
\end{equation}
The union between the proposed schemes, $\mathcal{S}^{\left(o\right)}\!=\!\mathcal{S}_\circ \bigcup \mathcal{S}_1 \bigcup \mathcal{S}_2\!=\!\mathcal{S}_\circ \bigcup \mathcal{S}_2$, can be achieved by using a switching optimization parameter to switch from one scheme to another in order to maximize the stability region of the SU given certain $\lambda_p$, $P_{\rm FA}$, $P_{\rm MD}$, and channels outage.
\section{Numerical results}\label{sec5}
\begin{figure}[h]
\center
  \includegraphics[width=0.9\columnwidth]{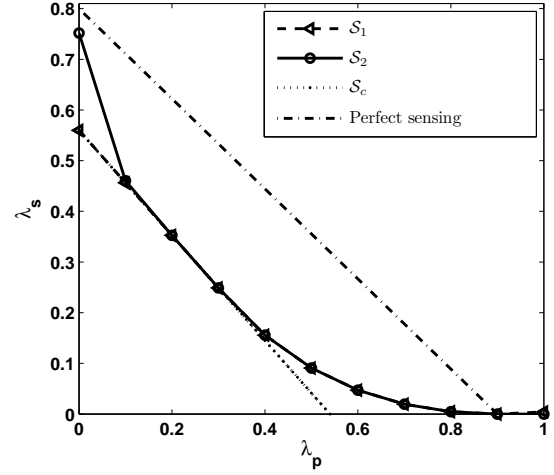}\\
 \caption{Stability region of the proposed system. The parameters used to generate the figure are: $P_{\rm MD}=0.3$, $P_{\rm FA}=0.2$, $\overline{P}_{p,pd}=0.9$, and $\overline{P}_{s,sd}=0.8$.}\label{fig8}
\end{figure}

\begin{figure}[h]
\center
  \includegraphics[width=0.9\columnwidth]{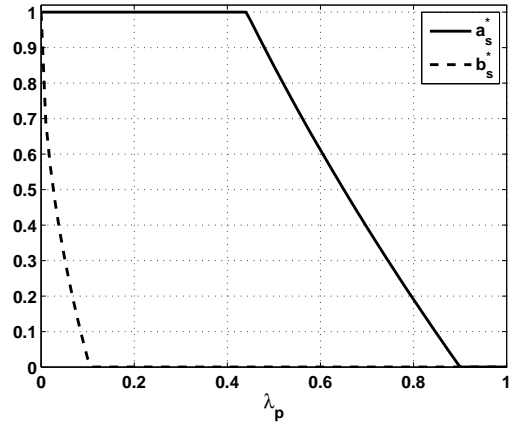}\\
 \caption{The optimal access probability for system $\mathcal{S}_2$ for specific $\tau$. The parameters used to generate the figure are: $P_{\rm MD}=0.3$, $P_{\rm FA}=0.2$, $\overline{P}_{p,pd}=0.9$, and $\overline{P}_{s,sd}=0.8$.}\label{fig800}
\end{figure}

\begin{figure}[h]
\center
  \includegraphics[width=0.9\columnwidth]{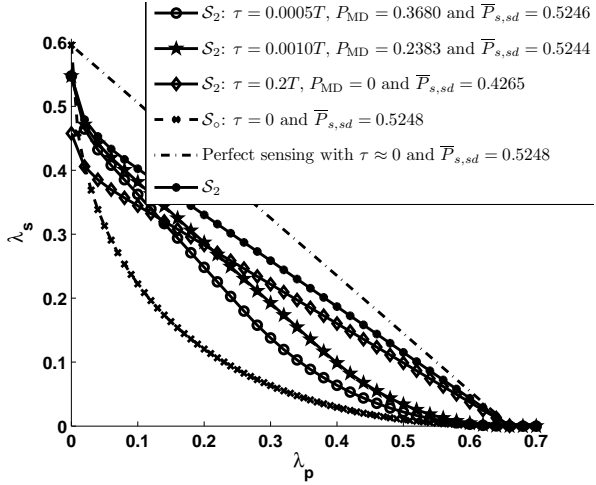}\\
  \caption{Stability region of the second proposed scheme, $\mathcal{S}_2$, as the sensing duration varies. The parameters used to generate the figure are: $P_{\rm FA}=0.2$ and $\overline{P}_{p,pd}=0.6609$.}\label{fig5}
\end{figure}

The maximum stable throughput of the considered schemes is shown in Figs. \ref{fig8} and \ref{fig5}. Fig.\ \ref{fig800} shows the optimal access probability of $\mathcal{S}_2$ as the mean arrival rate of the primary queue varies. It is noted that the optimal access probabilities are monotonic decreasing function of $\lambda_p$. This is because of the fact that as the arrival rate of the primary queue increases the probability of emptiness of the primary channel decreases, therefore the probability of silence periods vanishes and the SU will be unable to use the channel accordingly.

Fig \ref{fig5} shows the performance gain of $\mathcal{S}_2$ over $\mathcal{S}_c$ for different value of sensing duration. Also, the figure shows the stability region of $\mathcal{S}_2$ by taking the union over all possible sensing durations. Note that for small primary average arrival rate $\mathcal{S}_2$ with very small sensing duration overcomes $\mathcal{S}_\circ$ and $\mathcal{S}_2$ with long sensing duration. Also, $\mathcal{S}_\circ$ is better than $\mathcal{S}_2$ with long sensing duration because the SU does not waste $\tau$ seconds of transmission in sensing.

\begin{figure}[h]
\center
  \includegraphics[width=0.9\columnwidth]{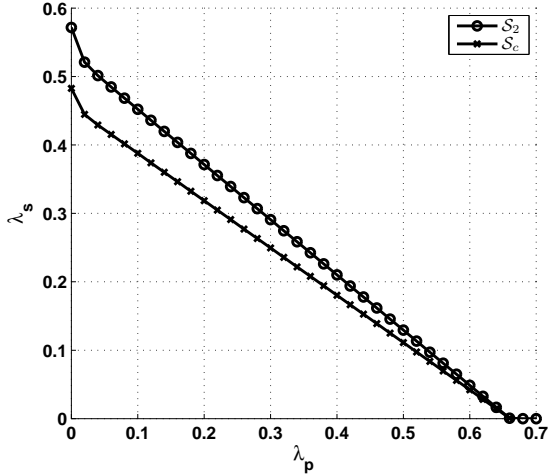}\\
  \caption{Stability region of the second proposed scheme, $\mathcal{S}_2$. The parameters used to generate the figure are: $P_{\rm FA}=0.1$ and $\overline{P}_{p,pd}=0.6609$.}\label{fig7}
\end{figure}

Fig. \ref{fig7} provides the solutions of the optimization problems (\ref{100}) and (\ref{200}). The figure compares between $\mathcal{S}_c$ and $\mathcal{S}_2$. It can be noted that the maximum stable throughput for the secondary node is monotonically decreasing function in $\lambda_p$, i.e., the mean service rate of the secondary node decreases to avoid violating stability conditions of the primary node (due to collisions). Note that this monotonic decreasing behavior of the mean service rate of the SU caused by the monotonic decreasing of the access probabilities of the SU with the primary queue arrival rate.
%\begin{figure}[h]
%\center
%  % Requires \usepackage{graphicx}
%  \includegraphics[width=0.9\columnwidth]{fig12}\\
%  \caption{Stability region of the second proposed scheme, $\mathcal{S}_2$, as the sensing duration varies. The parameters used to generate the figure are: $P_{\rm FA}=0.2$ and $\overline{P}_{p,pd}=0.6609$.}\label{fig6}
%\end{figure}

\section{Conclusion}\label{sec6}
In this work, we investigated the gains on the stability region of a SU randomly accesses the primary channel after making some sensing process. The results reveal the gains of the proposed schemes over the conventional sensing scheme and over the random access without sensing. The SU average service rate for the second proposed scheme with very small sensing duration can overcome sensing channel for long duration. We proposed variable sensing duration schemes where the SU optimizes over the optimal sensing time to achieve the maximum stable throughput for both primary and secondary queues.

The theoretical and numerical results show that pairs of misdetection and false alarm probabilities may exist such that sensing the primary channel for very small duration overcomes sensing it for large portion of the time slot. In addition, for certain average arrival rate to the primary queue pairs of misdetection and false alarm probabilities may exist such that the random access without sensing overcomes the random access with long sensing duration.

For very low misdetection and false alarm probabilities the proposed schemes are reduced to the conventional scheme, i.e., if sensing outcome is robust all schemes coincide. A switching optimization parameter can be used to switch from one scheme to another based on the maximum stable throughput.

 We proposed a method to estimate the mean arrival rate of the PU based on the primary feedback channel, i.e., ACKs and NACKs.

\section*{APPENDIX A}
In this Appendix, we provide the solution of the following optimization problem:
\begin{equation}
\begin{split}
    & \max_{x}\ \ \ \frac{a\ x +f}{c\ x - d}+K\ x, \\& {\rm s.t.} \   \ \ 0  \le  x  \le  \frac{d-w}{c},\  x \le 1
    \end{split}
\end{equation}
where $a,f,c,d,K$, and $w$ are \textbf{positive} constants. For the problem to be feasible $d$ should be greater than or equal $w$, i.e., $d\ge w$. The first derivative of the objective function gives:
\begin{eqnarray}
    \frac{a(c\ x-d)-c \ (a\ x +f)}{(c\ x -d)^2}+K=0
\end{eqnarray}
The second derivative is given by:
\begin{eqnarray}
    2c\frac{ad+cf}{(c\ x -d)^3}
\end{eqnarray}
Since $c\le d$, thus $c\ x -d \le 0$ and $x\in[0,1]$, therefore, the second derivative is always less than or equal zero. This implies that the function is \textbf{concave}.

After some mathematical manipulation for the first derivative
\begin{eqnarray}
(c\ x -d)^2=\frac{ad+cf}{K}
\end{eqnarray}
The roots of the quadratic equation are given by:
\begin{eqnarray}
x_1\!=\! \frac{d\!+\! \sqrt{\!\frac{ad\!+\!cf}{K}}}{c}, \ x_2\!=\! \frac{d\!-\! \sqrt{\!\frac{ad\!+\!cf}{K}}}{c}
\end{eqnarray}
One of the solutions is greater than the constraints which is $x_1>\frac{d-w}{c}$ (actually, it is greater than one because $c\le d$), thus $x^*=x_2$.
Including the constraints that $x\le 1$ and $x\ge 0$, thus the optimal value of $x$ is given by:
\begin{equation}
\begin{split}
x^*&\!=\! \max\biggr\{\min\biggr\{\! \frac{d\!-\! \sqrt{\!\frac{ad\!+\!cf}{K}}}{c},\frac{d\!-\!w}{c},1\biggr\},0\biggr\}
\end{split}
\end{equation}
with $d\ge w$.
\section*{APPENDIX B\\
NECESSARY AND SUFFICIENT CONDITIONS FOR STABILITY}
Given identical initial queue sizes for both the original and
dominant systems, secondary node queue in the dominant
system are never shorter than those in the original one. This is
true because in the dominant system, secondary and primary nodes suffer
from an increased collision probability, thus longer queues,
compared to the original one since secondary node always
has a packet to transmit, i.e., possibly a dummy packet. This
implies that secondary and primary queues empty faster in the original
system and therefore nodes see a lower probability of collision
as compared to the dominant system, and as a result will
have shorter queues. Consequently, stability conditions for the
dominant system are sufficient for the stability of the original
system.

To prove the necessary conditions, we follow an argument
similar to that used by \cite{rao1988stability} for ALOHA systems to prove the
\textbf{indistinguishability} of the dominant and original systems at
saturation. Consider the dominant system in which secondary
node transmits dummy packets. If the SU saturates in the original system, thus, no dummy packets are transmitted, then the original system and the dominant
system are indistinguishable. Thus, with a particular initial
condition, if the secondary queue in the dominant system never
empties with nonzero probability (i.e., it is unstable), then
secondary queue in the original system must be unstable as
well. This means that the boundary of the stability region of
the dominant system is also a boundary for the stability region
of the original system. Thus, conditions for stability of the dominant system are sufficient and necessary for the stability
of the original system.
\bibliographystyle{IEEEtran}
\bibliography{IEEEabrv,energy_bib}
\end{document}